# Some properties of hierarchic regular networks
# (a peep into fractal structures)

Gregory Surdutovich (1), Vladimir Gol'dshtein (2) and Gennady Koganov(2)

((1) Laser Physics Institute, Novosibirsk, Russia, (2)Ben Gurion University, Beer Sheva, Israel)

We give exact relations for certain types of the hierarchic fractal structures which characteristics in the limit of a large number of vertices acquire a simple elegant form. In the blatant distinction from regular networks of the "small world" (SW) topology [1], *regular fractal networks* manifests *the logarithmic* dependence of the characteristic path length on number of vertices $N$ which is typically just for random networks. Due to this fact, for the hierarchic networks there is no effect of an abrupt drop of the mean path length caused by the introduction into a regular structure only a few random edges. The overall drop of the characteristic path length under transition from the regular fractal to random network occurs to be not more than about factor three. The clustering coefficient of the hierarchic structures is equal zero and in a process of randomization it slightly grows up to the random networks value $k/N$, where $k$ is mean number of the nearest neighbors, in vivid contrast with the small world networks behavior. We considered also the problem of unification of two isolated SW and Caley tree networks in the limits of weak (through one point) and strong (many points) connection and deduced a specific algebra for the mean path length of the unified network in terms of the isolated networks path lengths. The sewing together of the tree-like networks may be performed in the one-bottom point or in the crown top to crown top manners. In such a way we come to the concept of bi-trees of two types having different topologies and essentially different characteristic mean path lengths. On this basis the periodical linear and ring clustered structures are proposed. Analytically and by the numerical simulation we demonstrated the difference between two possible probabilistic models of introducing randomness into a regular network: the model with substitution of a certain number of cut off links of the regular lattice by the equal number of edges stochastically distributed over the lattice (totally random model) and the artillery model, when one end of the broken link remains connected with the origin lattice. Finally, we discuss the problem of vulnerability of the hierarchic networks relatively of the process of their randomization.

**1. Introduction**. **SW effect**. After the seminal work by Watts and Strogatz [1] about the networks at the "interface" between regular and random ones the flux of papers devoted to such problems increased enormously. Authors had considered one special class of the one-dimensional lattices with nearest- and next-nearest neighbor connections that can be converted from ordered to a random by varying unique stochastic parameter $p$ (the normalized relative number of the random connections into a network) governing behavior (value) of two principal parameters of any network- the characteristic path length $L(p)$, defined as the average distance between two generic sites, and the clustering coefficient $C(p)$ describing cliquishness of a typical site's neighborhood. They demonstrated that such about regular networks with sparse admixture of the accidental links, named "small worlds" networks in analogy with similar phenomenon in social psychology [2], are in fact highly clustered like regular lattices (great $C$ ), yet have small characteristic path lengths $L$ like random graphs. The SW effect, i.e.the rapid drop in $L(p)$ after inclusion into the regular network of a few accidental links, may be easily explain qualitatively at example of the quadratic network. The review of the SW models is given in Ref. [3].

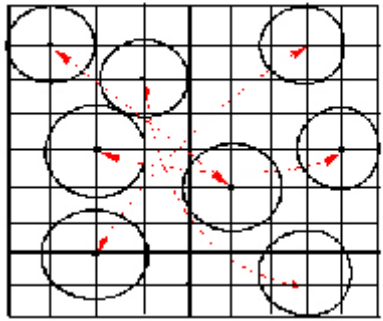

Fig. 1. After inclusion into a quadratic regular network with $N$ sites and $L_0 @ \sqrt{N}$ parameter of $m$ regular links by the same number of the random ones we obtain $m$ clustered regions with approximately $\frac{N}{m}$ sites in each with the regional parameter $L = \frac{L_0}{\sqrt{m}}$. Since these regions are connected by the random links it would be necessary to do only logarithmically small number of steps $\frac{\ln m}{\ln 2}$ to reach any region of the faceted network with the same regional parameter $L$. Therefore, in the logarithmic approximation, $m >> \ln m$, any points of the original network after inclusion of $m$ stochastic links will be separated by the distance $L = \frac{2L_0}{\sqrt{m}}$, i.e. a huge drop of the global parameter takes place.

2. **Hierarchic Caley tree like structures**. The understanding of the topological structure of networks and its connection with dynamical properties of a network became one of the principal directions of the investigations. Why is network anatomy so important to characterize? Because structure is always affects dynamical function of a network. Here we present some simple characteristics of hierarchic Caley tree-like networks, calculate their mean path lengths and derive the rules of their unification (sewing together).

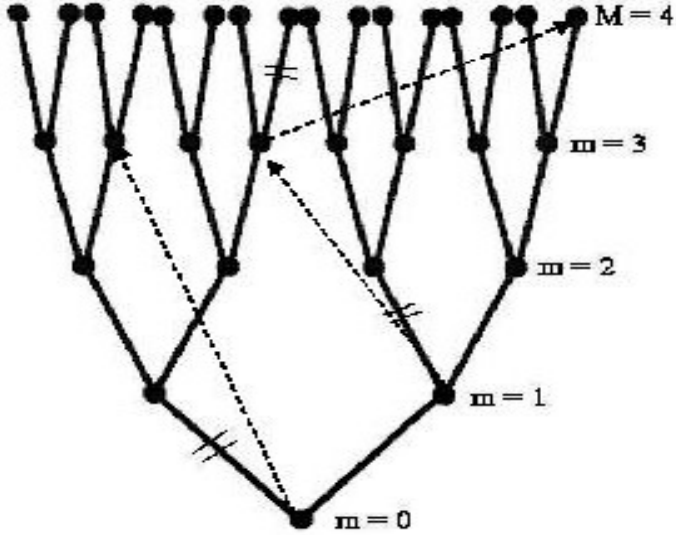

Fig.2.Here inclusion of the random links is shown by the dashed lines drawn from the bottom point of the broken link .It will be corresponds to the our "artillery model" (see further).

Let us consider tree-like fractile network with M layers (stories). Then there is the explicit relationship between total number of vertices N and mean path length $\mathbf{L}$ of a network. In terms of M we have

$$N = 2^{M+1} - 1 \qquad (1)$$

Bearing in mind applications to the diffusion process it is useful to divide all pathlengths into two groups –the horizontal path lengths $\mathbf{L_{hor}(m,m)}$ between vertices of the same horizontal level and $\mathbf{L_{vert}(m,m'), m \ne m'}$, between different levels. By the direct calculation one finds

$$\mathbf{L_{hor}(m,m) = (m-1)2^{2m+1} + 2^{m+1}} \qquad (2)$$

and for the total length $\mathbf{\tilde{L}_{hor}(M)}$ of all horizontal path lengths we obtain

$$\mathbf{\tilde{L}_{hor}(M) = \sum_{m=1}^{M} L_{hor}(m,m) = \frac{8}{3}\left[(M-1)2^{2M} + \frac{4}{3}(1 - 2^{2(M-1)})\right] + 4(2^M - 1)} \qquad (3)$$

The similar calculations for $\mathbf{\tilde{L}_{vert}}$ it is necessary take the total length of the directed (from top to bottom) path lengths between m and (m-k) levels

$$\mathbf{L_{vert}(m,m-k) = 2^m\{(2m-k-2)2^{m-k} + 2\}} \qquad (4)$$

and summarize them over all m and k levels

$$\mathbf{\tilde{L}_{vert} = 2\sum_{m=1}^{M}\sum_{k=1}^{m} 2^m\{(2m-k-2)2^{m-k} + 2\}} \qquad (5)$$

As a result we obtain

$$\mathbf{\tilde{L}_{vert} = \frac{4}{9}(12M - 28)2^{2M} + 4(M+3)2^M + \frac{4}{9}} \qquad (6)$$

To calculate the -mean path length of the network one should find a number of the horizontal, $\mathbf{S_{hor}}$, and vertical, $\mathbf{S_{vert}}$, pairs

$$\mathbf{S_{hor} = \frac{2}{3}(2^{M+1} - 1)(2^M - 1) = \frac{1}{2}S_{vert}} \qquad (7)$$

Note that according to Eqs.(3) and (6) in the main order $\tilde{L}_{hor}$ and $\tilde{L}_{vert}$ are equal to

$$\frac{8}{3}M2^{2M} \text{ and } \frac{16}{3}M2^{2M}, \qquad (8)$$

correspondingly. Therefore, mean path lengths

$$\overline{L}_{hor} = \frac{\tilde{L}_{hor}}{S_{hor}} \text{ and } \overline{L}_{vert} = \frac{\tilde{L}_{vert}}{S_{vert}} \qquad (9)$$

turn out be equal in this approximation ($M \gg 1$), when

$$S_{hor} = \frac{4}{3}2^{2M} \text{ and } S_{vert} = \frac{8}{3}2^{2M} \qquad (10)$$

So, in such zeroth approximation

$$\overline{L}_{hor} = \overline{L}_{vert} = \overline{L} = 2M \qquad (11)$$

Now we can write a general formula for $\overline{L}$ valid for any value of $M$. By combining Eqs.(3) and (6) one obtains

$$\tilde{L}_{total} = \tilde{L}_{hor} + \tilde{L}_{vert} = 8(M-2)2^{2M} + 4(M+4)2^M \qquad (12),$$

whereas, accordingly to Eq.(7)

$$S_{total} = S_{hor} + S_{vert} = 4 \times 2^{2M} - 6 \times 2^M + 2 \qquad (13)$$

Therefore, with an accuracy up to the exponential term

$$\overline{L} = (2M-4)(1 + 2 \times 2^{-M}) \qquad (14)$$

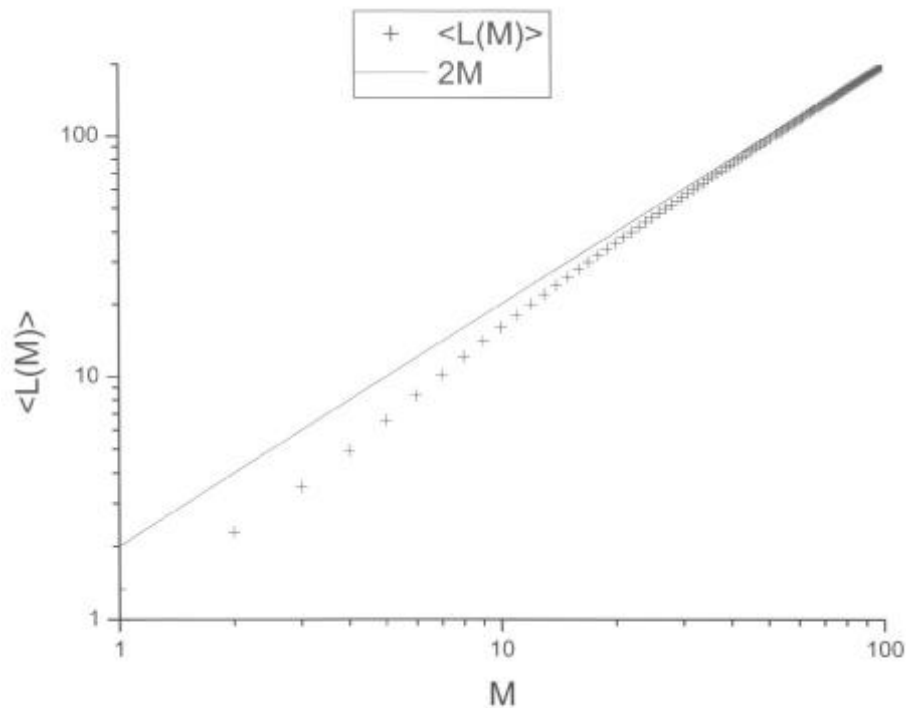

Fig. 3.The dependence of the global mean path length of a network as function of its upper level. Solid line corresponds to the asymptotic value of Eq.(14), whereas points correspond to the numerical simulations and Eqs.(12),(13).

For $M = 10$ the exponential correction term in Eq.(14) is lesser then 0.5% of the main order term. In the main order in number of vertices $N$ we have

$$\overline{L} = 2\frac{\ln N}{\ln 2} - 4 \qquad (15)$$

This result for regular fructal network one should compare with the corresponding mean path length $\overline{L}(p=1)$ of a totally stochastic network (stochastic parameter p=1 in notation of Ref. [1])

$$\overline{L}(p=1) = \frac{\ln N}{\ln k} = \frac{\ln N}{\ln 3} \qquad (16)$$

As a result, combining Eqs. (15) and (16) we obtain an estimate

$$\frac{L(p=0)}{L(p=1)} = \frac{2M-4}{M+1} \times \frac{\ln 3}{\ln 2} @ 1.584 \times \frac{2M-4}{M+1} \qquad (17)$$

Under any reasonable values of $M$ the decrease of mean path length in the process of stochastization of a network is not more than two-or threefold. This on the whole expected result that mean path `length of regular but highly hierarchic system only moderately changes under randomization of edges is connected with hierarchic topology of a network. The result is in the evident contradiction with abrupt drop of the mean path length under transition from regular uniform homogeneous to random networks. Moreover, the second peculiarity of our network is a fact that, in distinction from the SW models, the cliquishness parameter $C(p)$ *do not decrease* but *increases* from zero to the value of about $k/N$ under such process. Note, however, that zeroth value of the parameter $C(p=0)$ may be observed in some regular (not ring-like) nonhierarchic networks as well. The "topological meaning" of this result consists in the "topological proximity" of the hierarchic networks to the "central point-like" networks when all vertices are connected with one peculiar point (center) so that mean path length is small (lesser than two).

3**. The different tailor's variants of the Caley tree networks sewing together**. In Figs.4 and 5 are shown three-dimensional images of an isolated tree (Fig.4) and different tailor's variants of the possible structures (Fig.5).

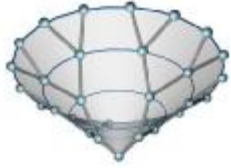
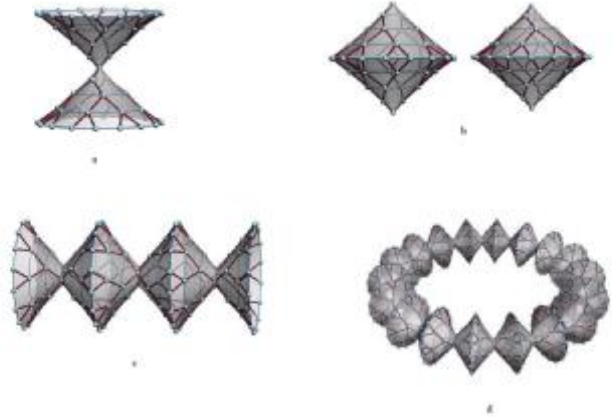

Fig. 4. An isolated Caley tree.

Fig. 5. a) Bi-tree structure obtained as the result of the point-to point sewing of two trees. b)Two variants of the crown to crown sewing: the left picture is the lobe to lobe sewing and right one is with half-period rotation of one of the trees. c) Periodical structure with two sews of trees at the different levels. d) Ring structure of sewn trees.

The hierarchic trees structures are of a special interest from the viewpoint of their randomization by use of the different probabilistic models. All tree-like structures, especially for small values of branching $(k - 1)$ are very sensitive to the manner of substitution of the broken regular links. We considered the $L(p)$-dependences of the be-trees structures by using two probabilistic models: with *a random choice* of both ends of a broken link (*random model*) and when one bottom end remains attached to the lattice (denoted as an *artillery model*) Here we give only such dependences for an isolated tree structure for random model (Fig. 6).

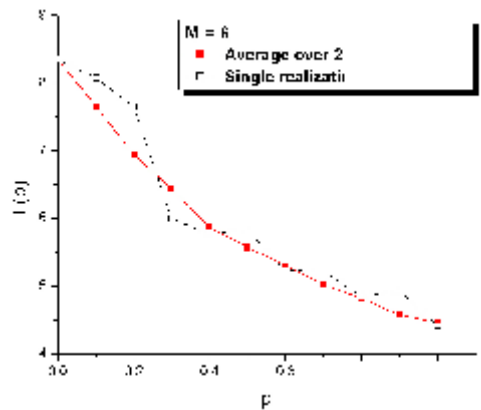

Fig. 6.The fluctuations in $L(p)$ dependence for an isolated $M$

For the probabilistic model with a random choice of both ends (artillery model) of the broken link the result is strongly depend on number of realizations (since there is great probability of disconnection of large parts of the network). Such a model with one fixed end corresponds to lesser probability choice and so the dependence $L(p)$ always diminishes more slowly (lies higher) than in previous case. For $p=1$ this corresponds as if effective two-fold diminution of number of random links so that in this model $L(p=1)$ is equal approximately to the value $L(p=0.5)$ in the previous model.

4. **Algebra of the tree-like networks' unifications.** An interesting property of such networks is the possibility to construct various combinations (unifications) of the tree-like fractile networks (conserving their topology and k-parameter!) connecting through only one (weak interaction) or through large number of links (strong interaction). In Fig.7 is shown the problem of two networks unification in the limit of a weak (one tie) coupling. Under such unification the total number of sites remains constant while by a change of average **k** due to the additional link may be neglected.

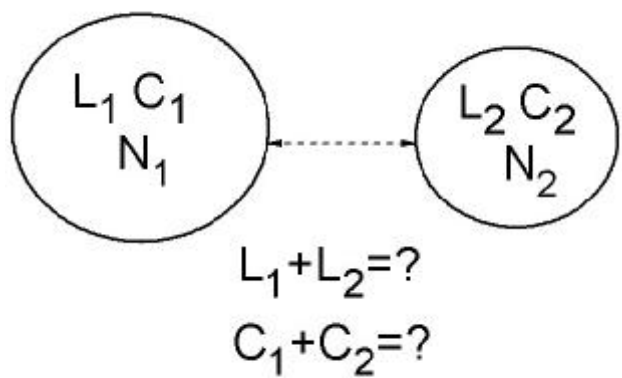

Fig. 7

It is evident that the total local parameter $C$ will change accordance with sizes of the networks

In the last case the number of connecting links may be comparable with total number of the networks vertices. Such a possibility is a specific property of the fractile (hierarchic) networks and does not exist for the conventional non-fractile networks. It is essential that in all these cases one can obtain exact analytical expressions for the global and local parameters and so separate the influence of network internal topologies and just networks "interaction" on final result of the unification. Firstly consider two simplest variants of unification-one-point and crown-crown sewing together.

a) *One-point sewing together* (Fig.4a.)

This case is a trivial. Two tree-like $\mathbf{M_1} \gg \mathbf{1}$ and $\mathbf{M_2} \gg \mathbf{1}$-levels networks with the parameters $\overline{\mathbf{L}}_1, \mathbf{C}_1$ and $\overline{\mathbf{L}}_2, \mathbf{C}_2$ after this type of unification will have the local parameter $\mathbf{C} = \mathbf{C}_1 = \mathbf{C}_2 = \mathbf{0}$ and global parameter $\overline{\mathbf{L}}$ given by

$$\overline{\mathbf{L}} = \frac{\mathbf{N}_1^2}{(\mathbf{N}_1+\mathbf{N}_2)^2}\mathbf{L}_1 + \frac{\mathbf{N}_2^2}{(\mathbf{N}_1+\mathbf{N}_2)^2}\mathbf{L}_2 + \frac{2\mathbf{N}_1\mathbf{N}_2(\mathbf{L}_{10}+\mathbf{L}_{20})}{(\mathbf{N}_1+\mathbf{N}_2)^2} \qquad (18)$$

where $\mathbf{L}_{10}$ and $\mathbf{L}_{20}$ are the mean pathlengths of networks connecting the original (bottom) point with al the rest vertices. To calculate them one should apply Eq.(4)

choosing $k = m$ to start all the trajectories from the ground point and the summarize over $m$ up to $m = M$. As the result for total length of the trajectories leading to the bottom point we obtain

$$L_{10} = \sum_{m=1}^{m=M} m 2^m = 2\{(M-1)2^M + 1\} \qquad (19)$$

The peculiar point here consists in absence (in distinction from Eq.(5)) of the factor **2**, since in the given case all the trajectories are directed towards to the bottom point (but not vice versa). If now divide Eq.(19) by a number of the trajectories $N - 1 = 2(2^M - 1)$, one obtains

$$\overline{L}_{10} = \frac{(M_1 - 1)2^{M_1} + 1}{2^{M_1} - 1} \qquad (20),$$

i.e. for $M_1 >> 1$ about twice lesser than $L_1$. One may write similar expression for $\overline{L}_{20}$.

In this approximation, $\overline{L}_{i0} = \frac{1}{2} L_i$ ($i = 1, 2$) Eq.(18) acquires the form

$$\overline{L} = \frac{N_1^2}{(N_1 + N_2)^2} L_1 + \frac{N_2^2}{(N_1 + N_2)^2} L_2 + \frac{N_1 N_2 (L_1 + L_2)}{(N_1 + N_2)^2} \qquad (21),$$

with twice lesser, in comparison with the similar expression for the conventional (non-fractal) networks, value of the last term.(see Ref.[4]). This distinction for the tree-like networks turns out to be very essential. Assuming a general case of non-identical networks let us introduce scale parameters $a = \frac{N_2}{N_1}$, $b = \frac{L_2}{L_1}$, describing the relative sizes and connectivities of the networks. Then Eq. (21) takes the form

$$\overline{L} = \{1 + \frac{a}{1+a}(b-1)\} L_1 = \{1 + \frac{a}{1+a}(b-1)\} \frac{1}{b} L_2 \qquad (22)$$

Therefore for any value of $b$ the mean pathlength of the unified network always remains between $L_1$ and $L_2$. Under $b = 1$ we have $\overline{L} = L_1 = L_2$, i.e. unification of the identical fractal networks do not lead to increase of the mean pathlength! This counterintuitive result differs from the corresponding formula of Ref.[2]

$$\overline{L} = \frac{3}{2} L_1 \qquad (22a)$$

for the conventional networks., when 50% of all pairs of the combined network have the doubled mean pathlength. The topological reason of such distinction for hierarchic networks consists in essential asymmetry (from the origin point to the top) of the tree-like network. The great majority of the vertices are disposed at the top of tree and are connected through the bottom levels by the trajectories of the "double-number of levels" length, whereas for the root-point of a tree the mean pathlength is comparable with height of a tree. Doubling of the inter-trees trajectories restores the value of the mean pathlength to the value of the mean pathlengths of the isolated networks.

   b) *Crown-crown sewing with overlapping of the sites* (Fig.4b).
   For the crown-crown variant of the sewing together one should calculate the total length of all trajectories between two networks. Firstly consider variant of sewing with "overlapping vertices" of the upper M levels of the networks. The total number sites of bi-tree sewed up from two M-level networks will be

$$N_{bi\text{-}tree} = 2(3 \times 2^{M-1} - 1) \qquad (23)$$

The bi-tree structure is symmetrical relatively of the intermediate M-layer and any inter-networks trajectory acquires length two under crossing of this layer.

$$\tilde{\tilde{L}}_{int} = (2^M - 1)\tilde{L}^1_{left\text{-}right}(M-1) + 2^M - (M+1) \qquad (24)$$

where $\tilde{L}^1_{left\text{-}right}(m) = (2M-1)2^{m+1} + M + 4$ is the total length of the trajectories going from any site of the $(M-1)^{th}$ level of the left network to all sites of the right network. To find the total length of all inter-networks trajectories in both directions one should summarize over all layers of the left networks

$$\tilde{\tilde{L}}_{inter\text{-}net} = \sum_{k=0}^{m} 2^k \{\tilde{L}^1(m) + m - k\} = (2^{m+1} - 1)[\tilde{L}^1(m) + m] - \sum_{k=0}^{m} k 2^k =$$
$$= (2M - 3)2^{2M+1} + (7 - M)2^{M+1} - 4(M+2) \qquad (25)$$

Then one should add to Eq.(25) the double summa of the intra-networks trajectories

$$\tilde{L}(M) = 8\{2(M-2)2^{2M} + (M+4)2^M\} \qquad (26)$$

with the exclusion of the double accounted horizontal trajectories of the $M^{th}$-level given by Eq.(2). As a result, one finds

$$\tilde{L}_{intra\text{-}net}(M) = (7M - 15)2^{2M+1} + (4M + 15)2^{M+1} \qquad (27)$$

By combining Eqs.(2),(25) and (26) one finds the total length $\tilde{\tilde{L}}_{total}$ of all directional trajectories of the unified network is

$$\tilde{\tilde{L}}_{total} = \tilde{\tilde{L}}_{inter\text{-}net} + 2\tilde{L}(M) - L_{hor}(M,M) = 9(M-2)2^{2M+1} + (3M+22)2^{M+1}$$
$$- 4M - 8 \qquad (28)$$

By dividing Eq.(28) by the total number of the pairs $N_{bi\text{-}tree}(N_{bi\text{-}tree} - 1) = 3(2^M - 1)(3 \times 2^M - 2)$ one finds the exact expression for the mean path length $\overline{L}_{bi\text{-}tree}$ of the unified bi-tree network. In the main order

$$\overline{L}_{bi\text{-}tree} = (2M - 4)(1 + 2 \times 2^{-M}) \qquad (29)$$

Comparing this result with Eq.(14) for the isolated tree network we see that in the limit of large M these expressions coincide up to to the exponential accuracy inclusive $\overline{L}_{bi\text{-}tree} = \overline{L}_{tree}$. It is very important result for construction of the hierarchic structures. Really, in contrast to the conventional case, under unification (sewing with overlapping) of the tree-like structures do not happen any increase of the total characteristic path length, while accordingly to Eq.(22a) for the unification process of similar conventional networks one and half increase occurs . Such distinction is connected not only with hierarchic topology of tree-like but with diminishing by about 25% , due to the overlapping under sewing together, of the total number of the networks poins. In an explicit form it becomes clear after comparison of this result with the variant of the unification of the networks *without overlapping procedure*, simply by connecting opposite sites of the upper layers of the networks by the additional links, while the total number of sites remains constant.

c) *The crown-crown sewing without overlapping of the sites*

Now we should re-calculate the total length of the inter-network trajectories, since now they connect not $(M-1)^{th}$ level trees through two-step layer but two $M^{th}$ level trees through additional one-step connection. Applying the previous procedure one should apply Eq.(25) with the substitution $M \to M+1$ and subtract from it the number of the inter-networks trajectories, since in comparison with the previous case each such trajectory became one step shorter, since in this variant the distance between two boundary $M$-layers is twice shorter of the distance between the $(M-1)^{th}$ layers of the previous structure. Finally, the total summa $\tilde{\tilde{L}}_{total}$ of the directional pathlengths is equal

$$\tilde{\tilde{L}}_{total} = \tilde{L}_{intra-net} + \tilde{L}_{inter-net} = 8(3M-4)2^{2M} + 48 \times 2^M - 4M - 14 \quad (30)$$

By dividing Eq.(30) by the number of the pairs $N_c(N_c - 1)$ in the unified network (where $N_c = 2(2^{M+1} - 1)^2$) one obtains for the mean pathlength in the main order

$$\overline{L}^{(c)}_{bi-tree} = 3M - 4 \quad (31)$$

For small values of $M$ this patlength may be about twice larger than in the variant b) while in the limit $M \gg 1$ we have

$$\overline{L}^{(c)}_{bi-tree} = \frac{3}{2}\overline{L}^{(b)}_{bi-tree} = \frac{3}{2}\overline{L}_{tree} \quad (32)$$

It is interesting that in this variant of the crown-crown of the networks sewing together (strong coupling) the result coincides with the unification of the conventional networks in the limit of a weak coupling (Eq.(22a)).

5. **Vulnerability and efficiency of the tree-like networks.** From a general viewpoint is evident that their vulnerability, i.e. sensitivity of the mean path length value to the "damage", cut-off of any certain link of a network, strongly depends on location of this cut-off: cut-offs in the upper level are of the least importance whereas the cut-off of the bottom link leads to disconnection of all the network. Formally it means infiniteness of the mean pathlength, i.e. maximum vulnerability of a tree-like networks in the conventional sense. An example of such approach to testing of sensitivity of a network to an occasional break of any link is given in Fig.9. for the electric power grid (COPEL) of state Parana'(Brazil), Ref.[4].

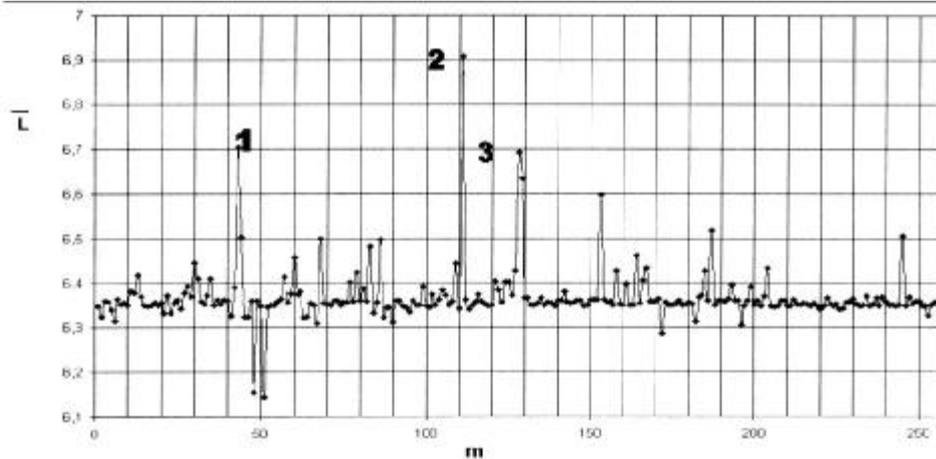

Fig. 8.The dependence of the global parameter $\overline{L}$ on break of any certain m-th link of the COPEL network. Among all 256 links of a network the connections denoted by 1, 2, 3 are especially vulnerable for function of the network.

We should note that for the hierarchic networks the other approach to the network's vulnerability may be more suitable. It consists in consideration of other (not L and C) the local and global efficiency parameters [5]. They are connected, in fact, with the inverse values of the path lengths and so suited for the consideration of the hierarchic networks where many configurations appeared in the process of randomization are related to the disconnected structures and so have inter-sites distances formally equal to infinity. Such change of the parameters resembles standard situation in the electrical networks when sometimes the conductivity variables are more suitable than the resistance ones. Now it seems fruitful to combine investigations about the static characteristics of the hierarchic structures with some new ideas about importance of a discrete description

The numerical simulations of the high-level trees and the sewn structures by use of a new much more power and rapid program are in the progress. One of us (G.S.) is grateful to the Mathematical Department of the Ben Gurion University for the hospitality.